\documentclass[prd,aps,preprint,amsmath,nofootinbib,amssymb,eqsecnum,showkeys,tightenlines]{revtex4}
\pdfoutput=1
\usepackage{verbatim,graphics,graphicx,color,slashed,textcomp}
\usepackage{ulem,soul,xcolor} 
\usepackage{hyperref}
\graphicspath{{figures/}}

\newcommand{\GeV}{\mathrm{GeV}}

\begin{document}
\setstcolor{red}

\title{
 Galactic center $\gamma$-ray excess in hidden sector DM models 
 with dark gauge symmetries: local $Z_{3}$ symmetry as an example}
\author{P. Ko and Yong Tang}
\affiliation{School of Physics, Korea Institute for Advanced Study,\\
 Seoul 130-722, Korea }
\date{\today}

\begin{abstract}
We show that hidden sector dark matter (DM) models with local dark gauge symmetries 
make a natural playground for the possible $\gamma$-ray excess from the galactic center (GC).  
We first  discuss in detail the GC $\gamma$-ray excess in a scalar dark matter (DM) model 
with local $Z_{3}$ symmetry which was recently proposed by the present authors. 
Within this model, scalar DM with mass $30-70$ GeV is allowed due to the newly-opened 
(semi-)annihilation channels of a DM pair into dark Higgs $\phi$ and/or dark photon $Z^{'}$ pair, and the $\gamma$-ray spectrum from the GC can be fit within this model.
Then we argue that the GC gamma ray excess can be easily accommodated within 
hidden sector dark matter models where DM is stabilized by local gauge symmetries, due to the presence of dark Higgs (and also dark photon for Abelian dark gauge symmetry).  
\end{abstract}
\maketitle

\section{Introduction}
Recently, it was claimed that analysis of data from the \textit{Fermi} Gamma-Ray Space Telescope has shown a possible excess of $\gamma$-ray from the Galactic Center (GC)~\cite{Daylan:2014rsa}, which was also reported in previous studies~\cite{Goodenough:2009gk, Hooper:2010mq, Boyarsky:2010dr, Hooper:2011ti, Abazajian:2012pn, Gordon:2013vta, Abazajian:2014fta}. 
Because galactic center is such a complex region, it is challenging to draw any definite conclusion 
at the moment. If we take this possible excess seriously, a cause for the excess 
is needed. Astrophysical explanation such as millisecond pulsars has been discussed~\cite{Yuan:2014rca}. However, it was claimed that signal from light dark matter annihilation is more favored~\cite{Daylan:2014rsa}.  
If one interprets this excess in terms of dark matter pair annihilation 
into a pair of standard model(SM) fermions, the annihilation cross section 
$\langle \sigma v \rangle_\textrm{ann}$ be at the order of the thermal relic cross section $10^{-26}\textrm{cm}^3$/s~\cite{Daylan:2014rsa}.
Both specific models and general frameworks have been investigated for this $\gamma$-ray 
signal~\cite{Lacroix:2014eea, Alves:2014yha, Berlin:2014tja, Agrawal:2014una, Izaguirre:2014vva, Ipek:2014gua, kpt, Boehm:2014bia, Abdullah:2014lla, Ghosh:2014pwa, Martin:2014sxa, Berlin:2014pya, Cline:2014dwa, Wang:2014elb, Agrawal:2014aoa, Han:2014nba, Cheung:2014lqa, Balazs:2014jla, Huang:2014cla, Fields:2014pia, Kong:2014haa, Arina:2014yna, Detmold:2014qqa, Bringmann:2014lpa, Cirelli:2014lwa} since then  
(see Refs.~\cite{Modak:2013jya, Boehm:2014hva, Hardy:2014dea} for some earlier discussion).
  
DM models for the GC $\gamma$-ray excess can be generally  divided into two categories. 
In the first category, DM annihilates directly into SM fermion pairs~\cite{Alves:2014yha, Berlin:2014tja, Agrawal:2014una, Izaguirre:2014vva, Ipek:2014gua, Ghosh:2014pwa, Wang:2014elb, Agrawal:2014aoa, Han:2014nba, Cheung:2014lqa,Balazs:2014jla, Huang:2014cla, Fields:2014pia}, 
which is however expected to be inconsistent with current constraints from collider and direct 
searches (see ref.~\cite{Basak:2014sza}, for example). 
In the second category, DM annihilates into two or more on-shell particles,  which in turn 
immediately decay into SM fermions~\cite{kpt, Boehm:2014bia, Abdullah:2014lla, Martin:2014sxa, Berlin:2014pya, Cline:2014dwa}.  A natural scenario of the second category can be easily 
realized in DM models with dark/hidden sectors and singlet portals (Higgs portal and kinetic mixing),
which is the main theme of this paper. In such models, one can easily evade the collider and direct search bounds, since DM particles interact with SM particles through singlet portals.

In this paper, we first discuss the GC $\gamma$-ray signal in scalar DM model with local $Z_3$ symmetry \cite{localz3} as a concrete example. In this model the dark sector has a local $U(1)_X$  gauge symmetry that is spontaneously broken into $Z_3$ subgroup, and two new particles 
(a dark Higgs and a massive dark photon) are introduced simultaneously due to underlying 
gauge symmetry. 
As discussed in Ref.~\cite{localz3}, 
the local $Z_3$ model has much  richer than and qualitatively different phenomenologies from 
the global $Z_3$ model~\cite{globalz3}, due to these two new particles~
\footnote{See Ref.~\cite{Aoki:2014cja} for effect of global $Z_3$ symmetry in a neutrino model.}. Semi-annihilation~\cite{Hambye:2008bq, D'Eramo:2010ep, Belanger:2012vp} breaks the tight correlation between thermal relic abundance and DM-nucleon scattering cross section, and allows 
scalar DM lighter than $125$GeV, in sharp contrast to the global $Z_3$ case~\cite{globalz3}.  
Also underlying local $Z_3$ symmetry guarantees the stability of scalar DM even in the presence 
of non-renormalizable higher dimensional operators. Furthermore, stabilizing DM particles via 
local gauge symmetries has a number of other virtues, as noticed in a series of recent works
~\cite{Baek:2013dwa,Baek:2013qwa,Ko:2014bka,localz3,Ko:2014uka}.

After the detailed discussions of the GC $\gamma$-ray signal in the local $Z_3$ scalar DM model, 
we generalize our finding to  general hidden sector DM models with local dark gauge symmetries. 
In this class of models, in addition to DM, there are new particles, namely dark gauge bosons and dark Higgs boson, whose interactions are completely determined by local dark gauge symmetry 
and renormalizability. Higgs portal interaction will be generically present in almost all the cases, 
and will thermalize DM relic density efficiently even for non-Abelian dark gauge symmetries (see 
Refs.~\cite{Hur:2007uz, Baek:2013dwa, Carone:2013wla, Hambye:2009fg, Holthausen:2013ota, 
Hur:2011sv, Heikinheimo:2013fta, Khoze:2014woa, Farzinnia:2013pga, Farzinnia:2014xia, 
Chiang:2013kqa,Boehm:2014bia} for some DM models based on non-Abelian groups). 
Higgs portal interaction also plays a very important role in direct and indirect detections of DM 
as well as in Higgs inflation~\cite{Ko:2014eia}.  

For Abelian dark gauge symmetry, there can be an additional portal from the $U(1)$ gauge kinetic mixing. Very often one just assumes a massive dark photon with this gauge kinetic mixing operator 
only, and consider various constraints from direct/indirect detections and thermal relic density altogether. However, couplings between DM and dark photon critically depend on the charge assignments to DM and dark Higgs fields (for example, compare the local $Z_3$ model
~\cite{localz3} and the local $Z_2$ model~\cite{localz2}), which is often overlooked in many works. 
Within local dark gauge theories, it is inconsistent to give mass to the dark gauge boson by hand, since it breaks  local dark gauge symmetry explicitly. It is important to introduce either dark Higgs  
field or some nonperturbative dynamical symmetry breaking mechanism to generate the dark 
gauge boson mass, while respecting local dark gauge symmetry and keeping all the allowed (renormalizable) operators. 
Otherwise, the resulting phenomenology could be misleading and sometimes even wrong, 
as shown in Ref.~\cite{Baek:2012se}.  

We note that DM models are also constrained by indirect searches. Stringent limits come from anti-proton and positron fluxes and radio signals~\cite{Kong:2014haa, Bringmann:2014lpa, Cirelli:2014lwa}. However, such constraints vary for different DM annihilation channels and 
also depend sensitively on various astrophysical factors as well: for example, the propagation 
parameters for the anti-proton flux, the local DM density for  the positron flux, and the 
DM density-profile at small radii $r < 5$pc for radio signals~\cite{Bringmann:2014lpa}, respectively. 
Currently, conservative limits still allow viable space for DM explanation of $\gamma$-ray excess. Due to the large astrophysical uncertainties, we shall not impose such constraints in our discussion,   but will show how the anti-proton flux depends on the propagation parameters as an example.

This paper is organized as follows. In Section.~\ref{sec:model}, we first introduce the scalar 
DM model with local $Z_3$ symmetry briefly, establishing the notations for later discussion. 
Then, in Section.~\ref{sec:gammaray}, we focus on the $\gamma$-ray spectrum from $Z_3$ scalar 
DM (semi-)annihilation from the GC and compare with the data. In Sec.~\ref{sec:lesson}, 
we generalize our finding to the general hidden sector DM models with dark gauge symmetries. Finally, we summarize the results in Sec.~\ref{sec:conclusion}.

\section{Scalar DM Model with Local $Z_3$ Symmetry}\label{sec:model}

In this section, we give a brief introduction of scalar DM model with local 
$Z_3$ symmetry~\cite{localz3}. The dark sector has a local $U(1)_{X}$ gauge symmetry that is spontaneously broken into $Z_{3}$ by the nonzero vacuum expectation value (VEV) of dark Higgs $\phi_X$. This can be realized with two complex scalar fields, 
\[
\phi_{X}\equiv\left(\phi_{R}+i\phi_{I}\right)/\sqrt{2}, \;
 X\equiv\left(X_{R}+iX_{I}\right)/\sqrt{2},
 \] 
with the $U(1)_{X}$ charges equal to $1$ and $1/3$, respectively. 
Then we can write down the most general renormalizable Lagrangian for the SM and dark 
sector fields, $\widetilde{X}_\mu, \phi_X$ and $X$: 
\begin{eqnarray}
{\cal L} & = & {\cal L}_{{\rm SM}}-\frac{1}{4}\tilde{X}_{\mu\nu}\tilde{X}^{\mu\nu}-\frac{1}{2}\sin\epsilon\tilde{X}_{\mu\nu}\tilde{B}^{\mu\nu}+D_{\mu}\phi_{X}^{\dagger}D^{\mu}\phi_{X}+D_{\mu}X^{\dagger}D^{\mu}X-V,\nonumber \\
V & = & -\mu_{H}^{2}H^{\dagger}H+\lambda_{H}\left(H^{\dagger}H\right)^{2}-\mu_{\phi}^{2}\phi_{X}^{\dagger}\phi_{X}+\lambda_{\phi}\left(\phi_{X}^{\dagger}\phi_{X}\right)^{2}+\mu_{X}^{2}X^{\dagger}X+\lambda_{X}\left(X^{\dagger}X\right)^{2}\nonumber \\
 &  & {}+\lambda_{\phi H}\phi_{X}^{\dagger}\phi_{X}H^{\dagger}H+\lambda_{\phi X}X^{\dagger}X\phi_{X}^{\dagger}\phi_{X}+\lambda_{HX}X^{\dagger}XH^{\dagger}H+
 \left( \lambda_{3}X^{3}\phi_{X}^{\dagger}+H.c. \right) ,  \label{eq:potential}
\end{eqnarray}
where the covariant derivative associated with the $U(1)_X$ gauge field $\tilde{X}^{\mu}$
is defined as $D_{\mu}\equiv\partial_{\mu}-i\tilde{g}_{X}Q_{X}\widetilde{X}_{\mu}$. 
The coupling $\lambda_{3}$ is chosen as real and positive, since one can always redefine 
the field $X$ and absorb the phase of $\lambda_3$.  

The vacuum phase relevant to our study should have the following structures:
\begin{eqnarray}
\langle H\rangle=\frac{1}{\sqrt{2}}\left(\begin{array}{c}
0\\
v_{h}
\end{array}\right),\;\langle\phi_{X}\rangle=\frac{v_{\phi}}{\sqrt{2}},\;\langle X\rangle=0,\label{eq:vacuumstate}
\end{eqnarray}
where only $H$ and $\phi_{X}$ have non-zero VEVs. Then the
electroweak symmetry will be broken into $U(1)_{\rm em}$.  The dark $U(1)_{X}$ gauge 
symmetry is broken into local $Z_3$, which stabilizes the scalar DM $X$.  
When expanding the scalar fields around Eq.~(\ref{eq:vacuumstate}), 
\begin{equation}
H\rightarrow\frac{v_{h}+h}{\sqrt{2}},\;\phi_{X}\rightarrow\frac{v_{\phi}+\phi}{\sqrt{2}},\; X\rightarrow\frac{x}{\sqrt{2}}e^{\mathrm{i}\theta}\textrm{ or }\frac{1}{\sqrt{2}}\left(X_{R}+iX_{I}\right),
\end{equation}
we find two scalar bosons $h$ and $\phi$ mix with each other through Higgs portal coupling 
$\lambda_{\phi H}$, resulting in two mass eigenstates $H_{1}$ and $H_{2}$ with
\begin{equation}
\left(\begin{array}{c}
H_{1}\\
H_{2}
\end{array}\right)=\left(\begin{array}{cc}
\cos{\alpha} & {}-\sin{\alpha}\\
\sin{\alpha} & \cos{\alpha}
\end{array}\right)\left(\begin{array}{c}
h\\
\phi
\end{array}\right),
\end{equation}
in terms of the mixing angle $\alpha$. We shall identify $H_1$ as the recent discovered Higgs 
boson with $M_{H_1}\simeq 125\GeV$ and treat $M_{H_2}$ as a free parameter. 
In principle, $H_2$ could be either heavier or lighter than $H_1$. 
However, we shall take $M_{H_2}\leq 80$ GeV in the following discussion,   
aiming at explaining the GC $\gamma$-ray excess in terms of 
$X \overline{X} \rightarrow H_2 H_2$ with scalar DM $X$ slightly heavier than $H_2$. 

After EW and dark gauge symmetry breaking, neutral gauge bosons can mix with each other, and the physical fields $( A_\mu , Z_\mu , Z_\mu^{'} )$ are defined as  
\begin{equation}
\left(\begin{array}{c}
\tilde{B}_{\mu}\\
\tilde{W}_{3\mu}\\
\tilde{X}_{\mu}
\end{array}\right)=\left(\begin{array}{ccc}
c_{\tilde{W}} & -\left(t_{\epsilon}s_{\xi}+s_{\tilde{W}}c_{\xi}\right) & s_{\tilde{W}}s_{\xi}-t_{\epsilon}c_{\xi}\\
s_{\tilde{W}} & c_{\tilde{W}}c_{\xi} & -c_{\tilde{W}}s_{\xi}\\
0 & s_{\xi}/c_{\epsilon} & c_{\xi}/c_{\epsilon}
\end{array}\right)\left(\begin{array}{c}
A_{\mu}\\
Z_{\mu}\\
Z'_{\mu}
\end{array}\right).\label{eq:mixing2}
\end{equation}
We have introduced new parameters: 
\begin{eqnarray}
 &  & c_{\tilde{W}}\equiv\cos\theta_{\tilde{W}}=\frac{g_{2}}{\sqrt{g_{1}^{2}+g_{2}^{2}}},\;\tan2\xi=-\frac{m_{\tilde{Z}}^{2}s_{\tilde{W}}\sin2\epsilon}{m_{\tilde{X}}^{2}-m_{\tilde{Z}}^{2}\left(c_{\epsilon}^{2}-s_{\epsilon}^{2}s_{\tilde{W}}^{2}\right)},\nonumber \\
 &  & t_{x}\equiv\tan{x},\; c_{x}\equiv\cos{x}\;\mathrm{and}\; s_{x}\equiv\sin{x}\;\mathrm{for}\; x=\epsilon,\xi,\nonumber \\
 &  &m_{\tilde{X}}^{2}=\hat{g}_{X}^{2}v_{\phi}^{2},\;\hat{g}_{X}=\tilde{g}_X/c_\epsilon,\; m_{\tilde{Z}}^{2}=\frac{1}{4}\left(g_{1}^{2}+g_{2}^{2}\right)v_{h}^{2}.
\end{eqnarray}
From Eq.~(\ref{eq:mixing2}) we can observe that SM particles charged under 
$SU(2)_{L}$ and/or $U(1)_{Y}$ now also have interactions with the dark photon $Z'_{\mu}$. And particles in the dark sector can  interact with $Z_{\mu}$ as well, 
due to the kinetic mixing beween $\tilde{B}_\mu$ and $\tilde{X}_\mu$. 
The physical masses for four gauge bosons in our model are given by 
\begin{align}
m_{A}^{2} =  0,\;   & m_{W}^{2}  =  m_{\tilde{W}}^{2}=\frac{1}{4}g_{2}^{2}v_{h}^{2},  \\
m_{Z}^{2} =  m_{\tilde{Z}}^{2}\left(1+s_{\tilde{W}}t_{\xi}t_{\epsilon}\right),\;
& m_{Z'}^{2}  =  \frac{m_{\tilde{X}}^{2}}{c_{\epsilon}^{2}\left(1+s_{\tilde{W}}t_{\xi}t_{\epsilon}\right)}
\end{align}
at tree level. 

In the scalar DM models with global $Z_3$ symmetry \cite{globalz3}, light dark matter ($m_\textrm{DM}\lesssim 125$ GeV) is generally excluded by LUX direct search experiment except 
for the resonance regime. 
On the other hand, in the scalar DM models with local $Z_3$ gauge 
symmetry, such light dark matter is still allowed due to the newly open annihilation channels
(see Ref.~\cite{localz3} for details). In this paper, we shall focus only on the indirect 
signatures in terms of $\gamma$-ray, anti-proton and positron fluxes within the local $Z_3$ 
scalar DM model. 

\section{$\gamma$-ray from DM (semi-)annihilation}\label{sec:gammaray}
\begin{figure}[t]
\includegraphics[width=0.7\textwidth, height=0.16\textwidth]{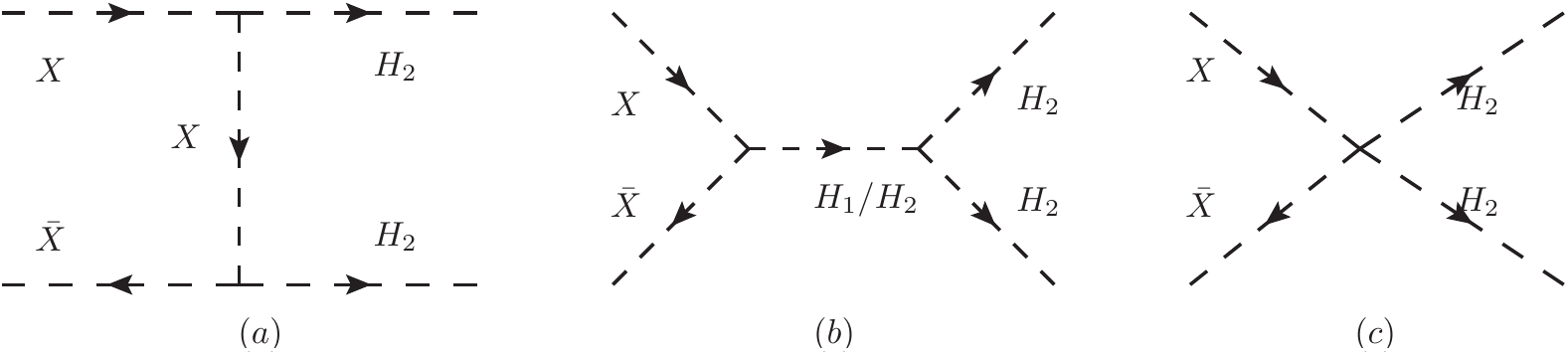}
\includegraphics[width=0.7\textwidth, height=0.16\textwidth]{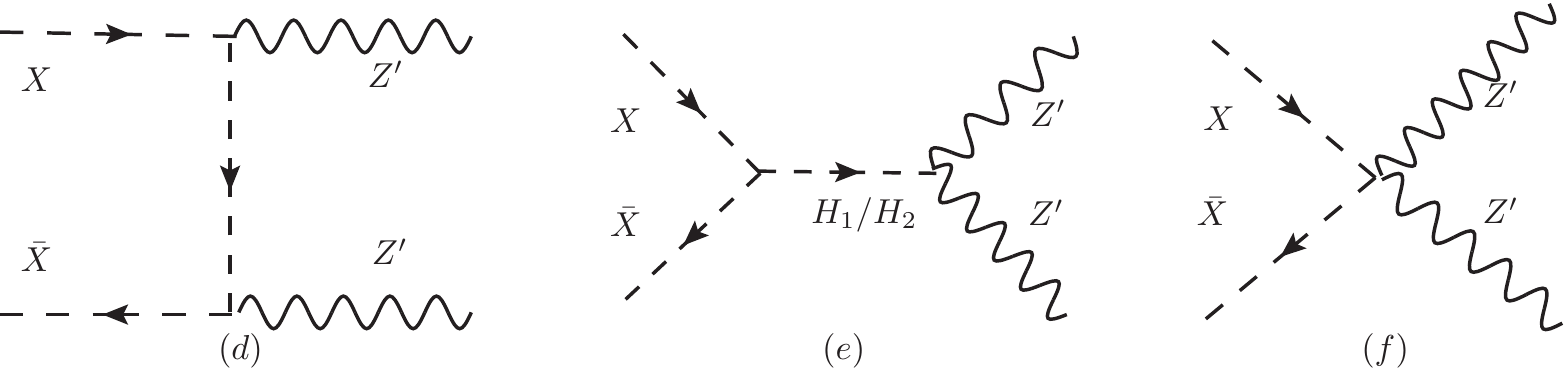} 
\caption{Feynman diagrams for $X\bar{X}$ annihilation into $H_2$ and $Z'$.
\label{fig:annihilation}}
\end{figure}
\begin{figure}[t]
\includegraphics[width=0.9\textwidth, height=0.16\textwidth]{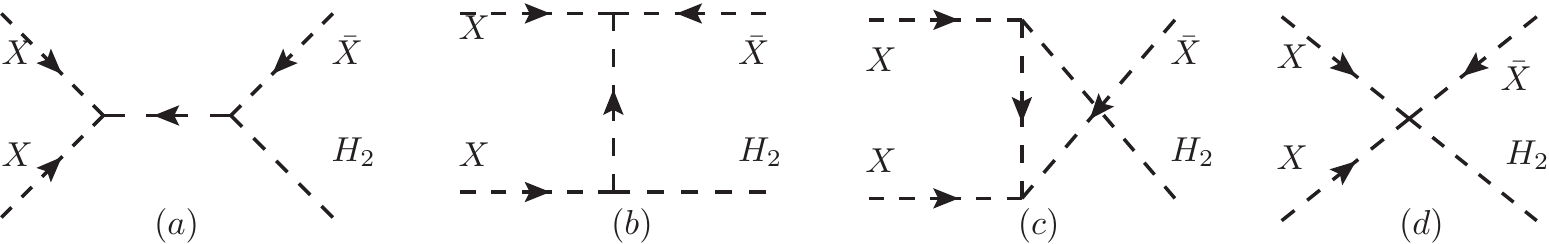}
\includegraphics[width=0.7\textwidth, height=0.16\textwidth]{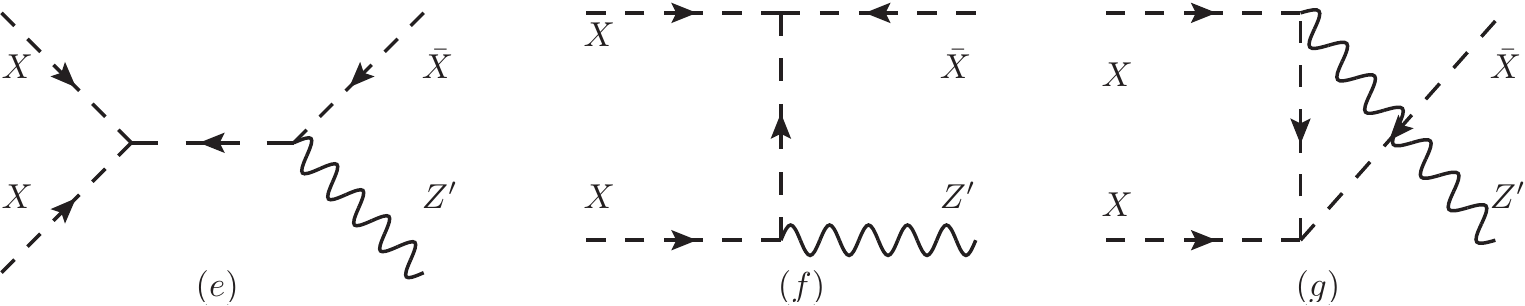} 
\caption{Feynman diagrams for $XX$ semi-annihilation into $H_2$ and $Z'$. 
\label{fig:semi-annihilation}}
\end{figure}

In the section, we shall discuss the $\gamma$-ray spectrum from dark matter (semi-)annihilation 
in the scalar DM model with local $Z_3$ symmetry. We shall focus on the channels shown in 
Figs.~\ref{fig:annihilation} and \ref{fig:semi-annihilation}, where  $H_2$s and $Z'$s in the final 
states decay into SM particles. DM pair annihilations directly into a pair of SM particles such as 
\[
 X \bar{X} \rightarrow (Z^{'*}  ~{\rm or}~  H_2^{'*}) \rightarrow \bar{f} f , 
\]  
are suppressed by the small mixing parameters, $\alpha$ and $\epsilon$. 
In the parameter regions we are interested in, we can take $\alpha$ and $\epsilon$ to be 
smaller than $10^{-4}$, which is definitely allowed by direct searches so far. 
For simplicity, we also assume vanishing $\lambda_{\phi H}$ and $\lambda_{HX}$. 
Non-vanishing $\lambda_{\phi H}$ and $\lambda_{HX}$ would not change  qualitatively our discussion.  Both parameters are constrained by DM direct searches, the  invisible branching 
ratio of Higgs boson, and collider bounds, discussion of which is beyond the scope of this paper.  

The $\gamma$-ray flux from self-conjugate DM (semi-)annihilation is determined by particle physics factors, $\langle \sigma v\rangle_\textrm{ann}$ and $d N_\gamma / dE_\gamma$, and astrophysical 
factor, DM density profile $\rho$:  
\begin{equation}\label{eq:flux}
\frac{d^2 \Phi}{dE_\gamma d\Omega}=\frac{1}{8\pi}\frac{\langle \sigma v\rangle_\textrm{ann}}{ M_{\textrm{DM}}^2}\frac{d N_\gamma }{dE_\gamma}\int_0^\infty dr \rho ^2 \left(r', \theta \right). 
\end{equation}
Here $r'=\sqrt{r_\odot^2 + r^2 -2r_\odot  r \cos \theta}$, where $r$ is the distance to earth 
from the DM annihilation point,  $r_\odot\simeq 8.5$kpc and $\theta$ is the observation angle between the line of sight and the center of Milky Way, respectively. An extra factor $1/2$ 
has to be included when $X$ annihilates with its anti-particle $\bar{X}$,
which is relevant to the local $Z_3$ scalar DM model.

\begin{figure}[t]
\includegraphics[width=0.6\textwidth, height=0.5\textwidth]{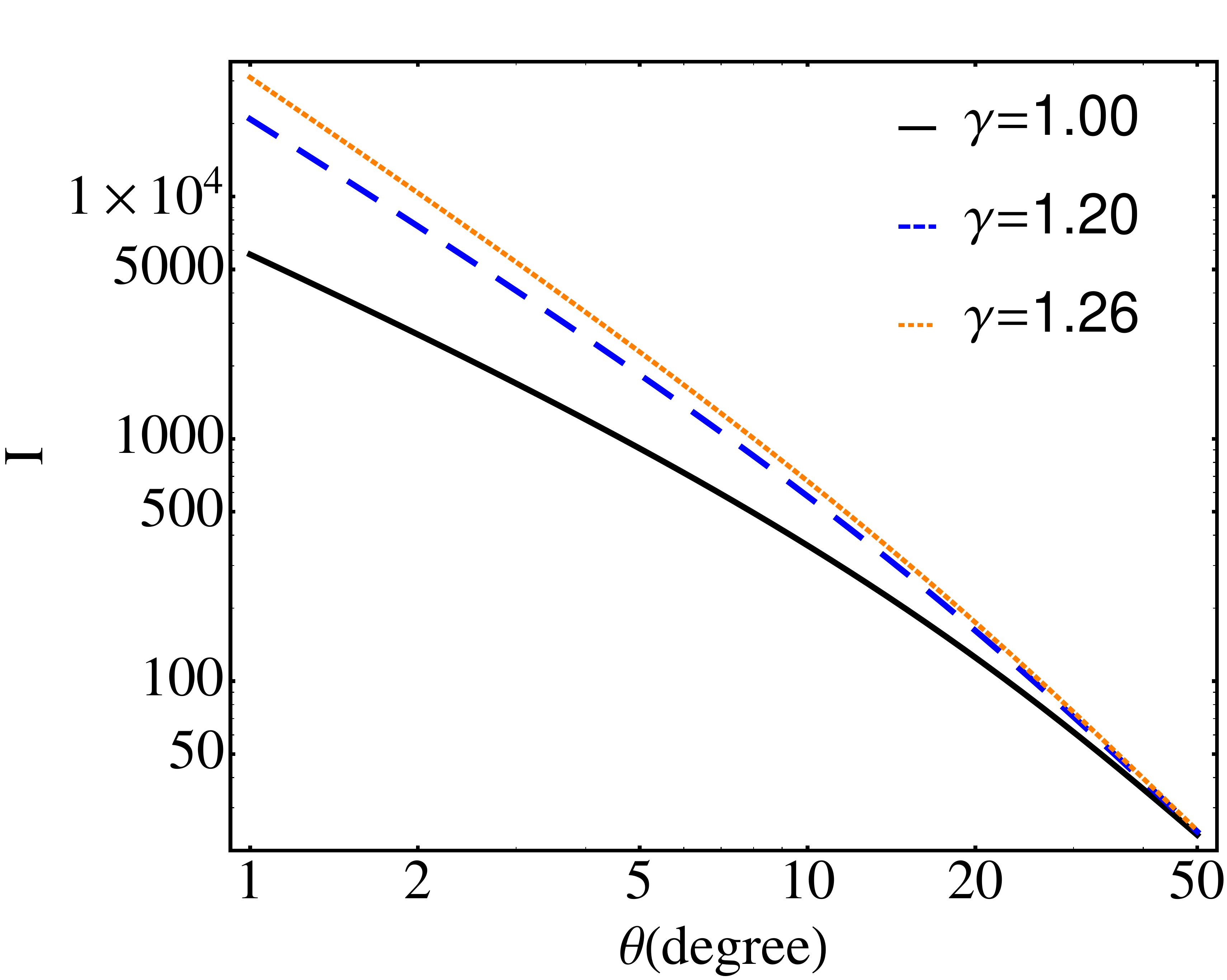} 
\caption{Dependence of $I$ on the index $\gamma$ and the angle $\theta$. 
$\gamma=1$ is for standard NFW density profile. The purpose of this plot is to show that there 
is a large uncertainty in dark matter density $\rho$ near the galaxy center.  
\label{fig:Ifactor}}
\end{figure}

We use the generalized NFW profile~\cite{Navarro:1995iw} for DM density, which is 
parametrized as   
\begin{equation}\label{eq:haloprofile}
\rho\left(r\right)=\rho_\odot \left[\frac{r_\odot}{r}\right]^\gamma \left[\frac{1+r_\odot/r_c}{1+r/r_c}\right]^{3-\gamma},
\end{equation}
with $r_c\simeq 20$kpc and $\rho_\odot \simeq 0.3\textrm{GeV}/\textrm{cm}^3$. 
Defining a dimensionless function $I$,
\begin{equation}
I\equiv \int_0^\infty  \frac{dr}{r_\odot} \left(\rho\left(r', \theta \right)/\rho_\odot\right)^2,  
\end{equation}
we show how $I$ depends on the power index $\gamma$ in Fig.~\ref{fig:Ifactor}. 
Because of the large uncertainty of $\rho$, it is not so meaningful to quantify the exact value for 
$\langle \sigma v\rangle _{\textrm{ann}}$ in Eq.~(\ref{eq:flux}), as long as it is at the order of $10^{-26}\textrm{cm}^3/$s.
We can roughly estimate $\langle \sigma v\rangle _{\textrm{ann}}$ in our model by fixing index $\gamma=1.26$. For self-conjugate DM annihilation, it was shown in~\cite{Daylan:2014rsa} that 
$\langle \sigma v\rangle _{\textrm{ann}}\simeq 1.7(1.1)\times 10^{-26}\textrm{cm}^3/$s can 
fit the $\gamma$-ray spectrum well for $b\bar{b}$(democratic) channel. 
It is then straightforward to get $\langle \sigma v\rangle _{\textrm{ann}}\simeq 6.8(4.4)\times 10^{-26}\textrm{cm}^3/$s for complex scalar or Dirac fermion DM that annihilates first into two on-shell particles $H_2$ (or $Z'$), which in turn decay into the SM particles through mixings. 
In the following discussion, we shall fix the index $\gamma=1.26$ and treat $\langle \sigma v\rangle _{\textrm{ann}}$ as a free parameter \footnote{One can also fix 
$\langle \sigma v\rangle _{\textrm{ann}}$, and treat the index $\gamma$ as a free parameter.}. 
This also means that at this stage we do not consider thermal relic abundance precisely. 

Due to the mixings among neutral gauge bosons, $Z'$ can decay into SM fermion pairs with branching ratios depending on flavours. For instance, the couplings for $Z_{\mu}'\bar{f}_R\gamma^\mu f_R$ and $Z_{\mu}'\bar{f}_L\gamma^\mu f_L$ are 
proportional to 
\begin{eqnarray}
   &g_1 (s_{\tilde{W}}s_{\xi}-t_{\epsilon}c_{\xi}) Y, \\
   &g_1 (s_{\tilde{W}}s_{\xi}-t_{\epsilon}c_{\xi}) Y - g_2 c_{\tilde{W}}s_{\xi} T_3,
\end{eqnarray}
respectively.  Here $Y$ is the $U(1)_Y$ hypercharge and $T_3$ is related to $SU(2)_L$. 
On the other hand, the dark Higgs boson $H_2$ couples to the standard model fermions 
in the same way as the SM Higgs boson does, except that the couplings are rescaled by the 
mixing angle factor $\sin \alpha$. Therefore the branching ratios of $H_2$ into the SM particles 
will be similar to those of the SM Higgs boson.

The shape of $\gamma$-ray spectrum $d N_\gamma / dE_\gamma$ depends on the 
mass parameters of the involved particles, $m_{X}$, $m_{H_2}$ and $m_{Z'}$, and the relative contribution of each annihilation channel in Figs.~\ref{fig:annihilation} and \ref{fig:semi-annihilation}, 
which in turn depend on the parameters $\lambda_3,\lambda_{\phi X}$ and $g_X$. 
Here for the purpose of simple illustration, we discuss two extreme cases:  either $100\%$ to 
$H_2$ or $Z'$ in the final states of DM (semi-)annihilation. A large parameter space exists 
between these two extreme cases. Dedicated analysis would require multi-dimensional 
$\chi^2$ fit, which is beyond the scope of this paper.  
We shall use \texttt{micrOMEGAs-3}\cite{micromegas3} for our numerical calculations and for 
generation of the $\gamma$-ray, anti-proton and positron spectra.

\begin{figure}[t]
\includegraphics[width=0.47\textwidth, height=0.42\textwidth]{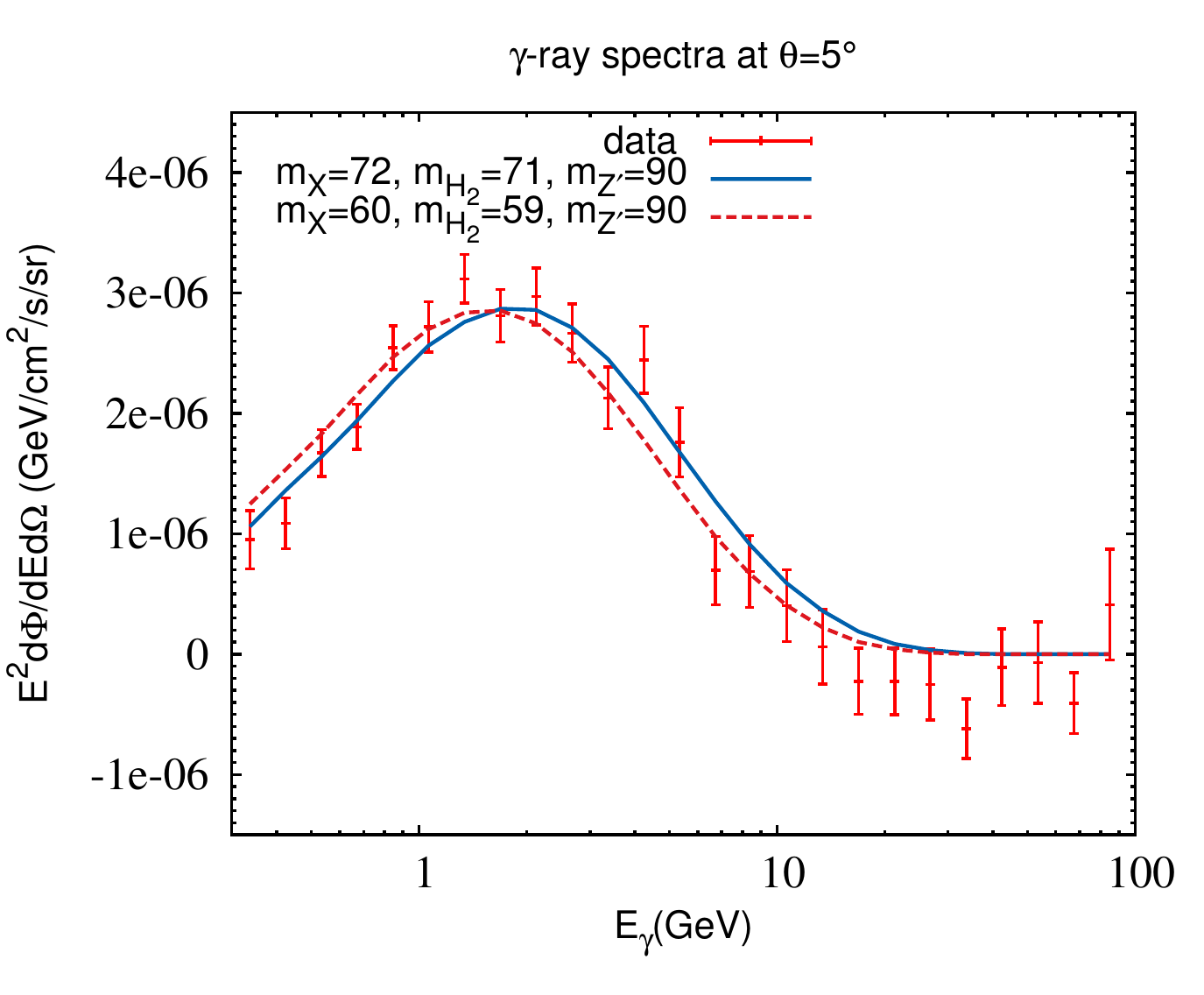}
\includegraphics[width=0.47\textwidth, height=0.42\textwidth]{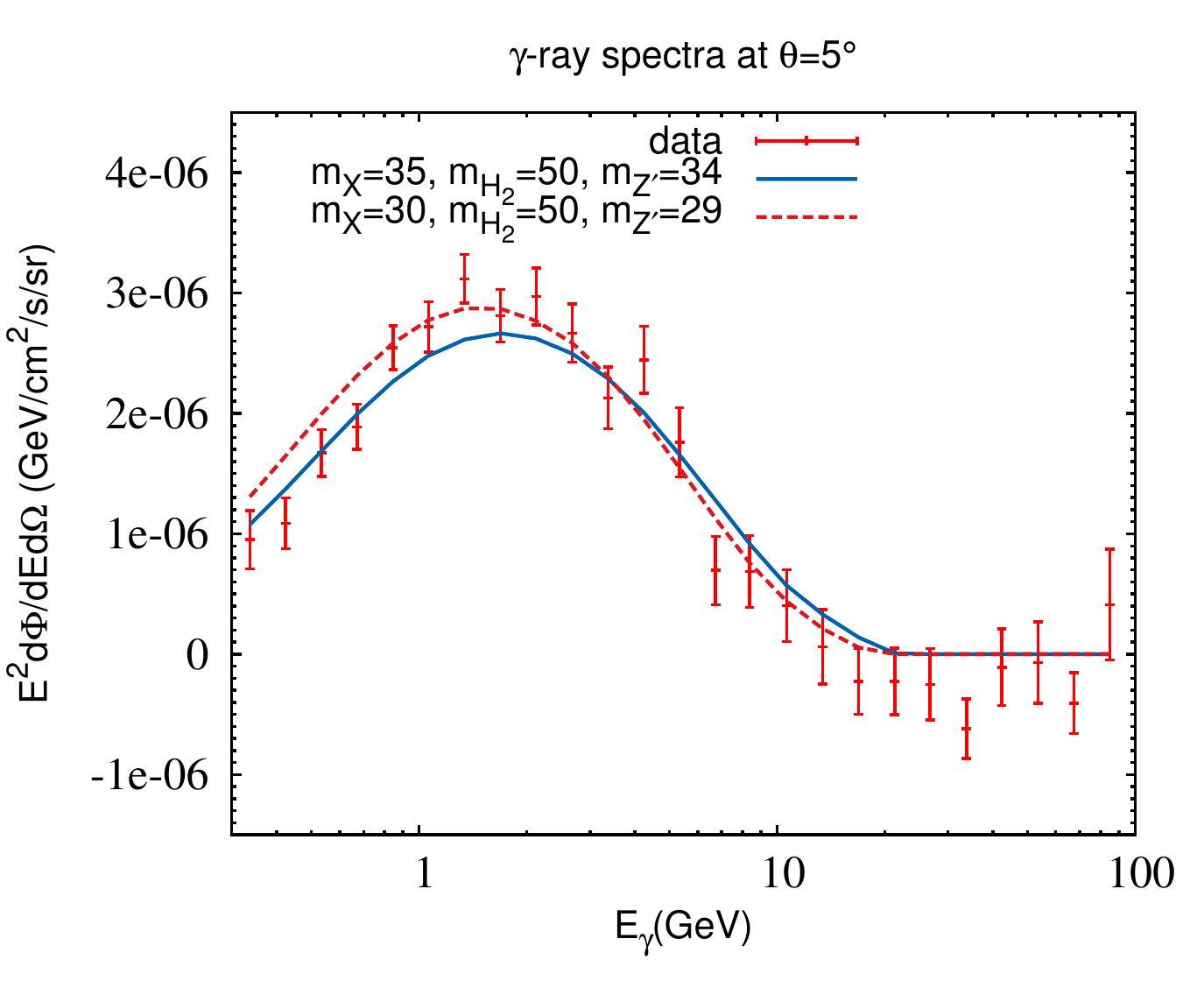} 
\caption{$\gamma$-ray spectra from dark matter (semi-)annihilation with $H_2$(left) and $Z'$(right) as final states. In each case, mass of $H_2$ or $Z'$ is chosen to be close to $m_X$ to avoid large lorentz boost. Masses are in GeV unit. Data points at $\theta=5$ degree are extracted from \cite{Daylan:2014rsa}.
\label{fig:gammaspectrum}}
\end{figure}

We show the $\gamma$-ray spectra in Fig.~\ref{fig:gammaspectrum} from $H_2$ (the left panel) 
and $Z'$ (the right panel).   Since we are discussing two extreme cases, we choose the mass of 
$H_2/Z'$ to be close to $m_X$. It is seen that $m_X$ is around $70$GeV for $H_2$ case, 
while $m_X\sim 30$GeV for $Z'$ case.
This is due to the fact that the dark Higgs boson $H_2$ with mass $\sim 70$GeV mainly 
decays into $b\bar{b}$ which give a softer $\gamma$-ray spectrum,  than the 
dark photon $Z'$ that would decay into charged fermion pairs.
For $30\textrm{GeV}\lesssim m_X \lesssim 70$GeV, we can adjust the relative contributions 
from $H_2/Z'$-channels, and it is anticipated that this can be  easily achieved by varying 
$m_{H_2}, m_{Z'}, \lambda_3,\lambda_{\phi X}$ and $g_X$.

\begin{figure}[t]
\includegraphics[width=0.47\textwidth, height=0.42\textwidth]{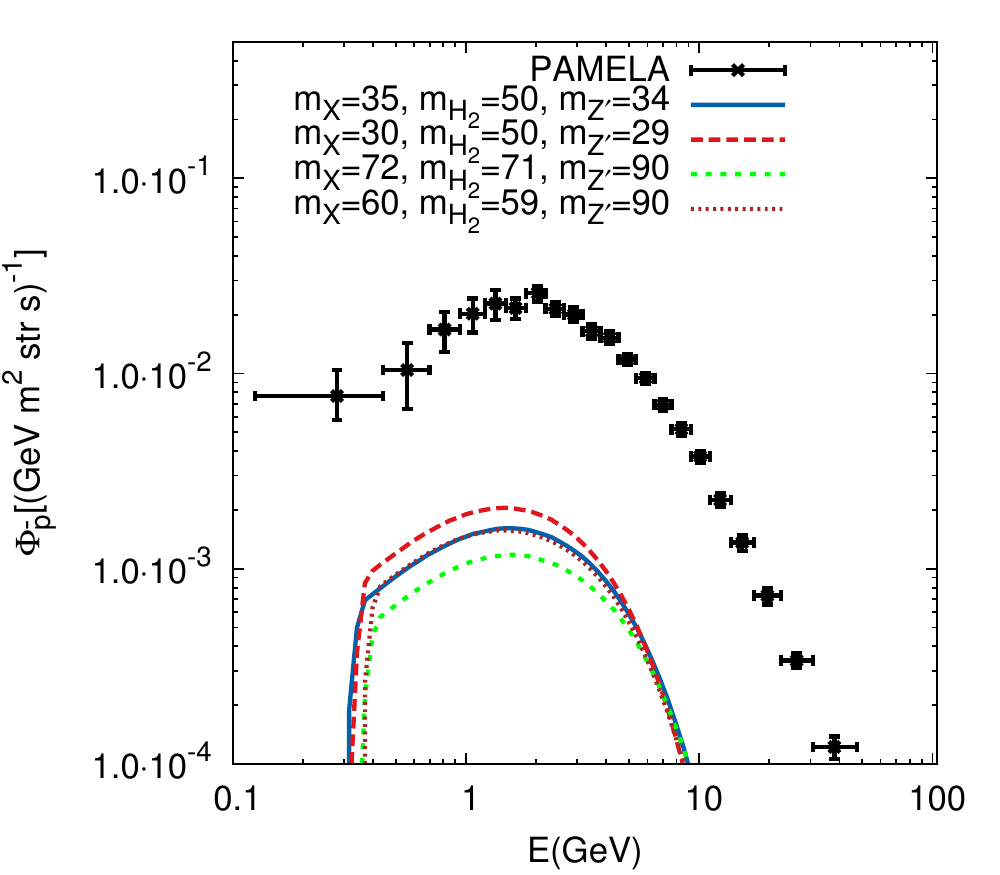}
\includegraphics[width=0.47\textwidth, height=0.42\textwidth]{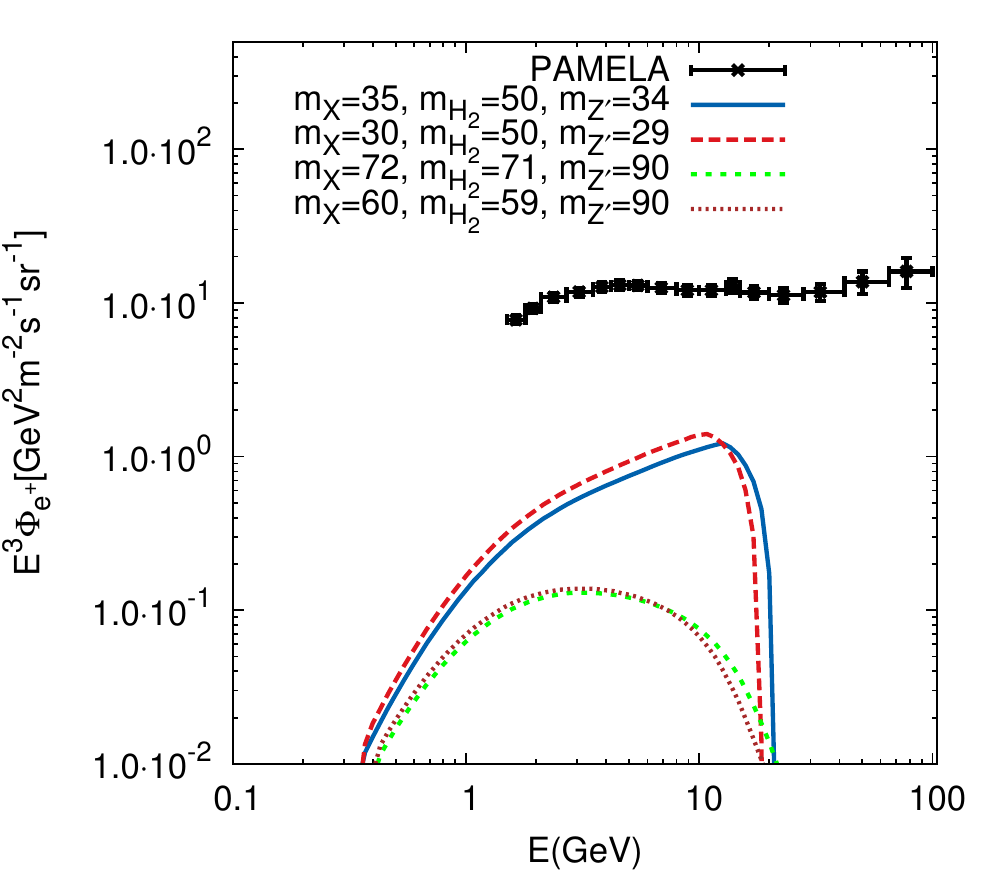} 
\caption{$\bar{p}$ and $e^+$ spectra from dark matter (semi-)annihilation with $H_2$(left) and $Z'$(right) as final states. In each case, mass of $H_2$ or $Z'$ is chosen to be close to $m_X$ to avoid large lorentz boost. Masses are in GeV unit. $\langle \sigma v \rangle _{\textrm{ann}}\simeq 6.8(4.4)\times 10^{-26}\textrm{cm}^3/$s for $H_2(Z')$ final states are assumed. Data point are taken from~\cite{ref:antiproton} for anti-proton and~\cite{ref:positron} for positron fluxes, using the database~\cite{crdb}. 
\label{fig:antiparticle}}
\end{figure}

For the same sets of parameters which accommodate the $\gamma$ ray excess from the GC,
we also studied $\bar{p}$ and $e^+$ spectra, as shown in Fig.~\ref{fig:antiparticle}, and 
checked if our choices of parameters are compatible with these cosmic ray spectra or not. 
Charged particles generated by DM (semi-)annihilation will propagate to the earth, 
subject to diffusion, synchrotron radiation and inverse Compton scattering. 
We use the \texttt{micrOMEGAs-3}~\cite{micromegas3} to calculate their spectra, 
with the MIN model being used for anti-proton propagation. 
As we can see from Fig.~\ref{fig:antiparticle}, the 4 sets of parameters give almost the same 
predictions for the $\bar{p}$ flux, whereas the resulting $e^+$ fluxes can differ by an 
order of magnitude, depending on whether DM pair annihilates into $H_2 H_2$ or $Z' Z'$. 
This is because the decay of $Z'$ can produce much harder $e^+$s.

In Fig.~\ref{fig:antiparticle} we show the signals from DM (semi-)annihilation only. 
After being added to the astrophysical background, these fluxes could be compared with the data from PAMELA~\cite{ref:antiproton,ref:positron}. The constraints from $\bar{p}$ and $e^+$ can provide important and  complementary information for DM models explaining $\gamma$-ray excess. 
It should be pointed out that potentially stringent constraints from indirect detections of cosmic 
rays depend sensitively on astrophysical parameters involved in the calculations of cosmic ray 
production and propagation. 

The propagation equation that describes the evolution of
energy distribution for charged particle $a$ is given by ~\cite{micromegas24} 
\begin{equation}
\label{eq:propa}
\frac{\partial}{\partial z} \left(V_C\psi_a\right)
- {\bf \nabla}\cdot\left( K(E) {\bf \nabla} \psi_a \right)-
\frac{\partial}{\partial E} \left( b(E) \psi_a \right) =Q_a({\bf x},E),
\end{equation}
where $\psi_a=dn/dE$ is the number density of particle $a$ per unit volume and energy. 
$Q_a$ is the source term for particle $a$ originating from dark matter annihilation. 
The function $K$ is the space diffusion coefficient which depends on the energy $E$:  
\begin{equation}
K(E)=K_0 \beta(E) \left({\cal R}/1\;{\rm GV}\right)^\delta.
\end{equation}
Here $\beta$ is the particle velocity, ${\cal R}=p/q$ is its rigidity($p$ is the momentum and $q$ is the charge) , and $b(E)$ is the energy loss rate. 

\begin{figure}[t]
\includegraphics[width=0.32\textwidth, height=0.32\textwidth]{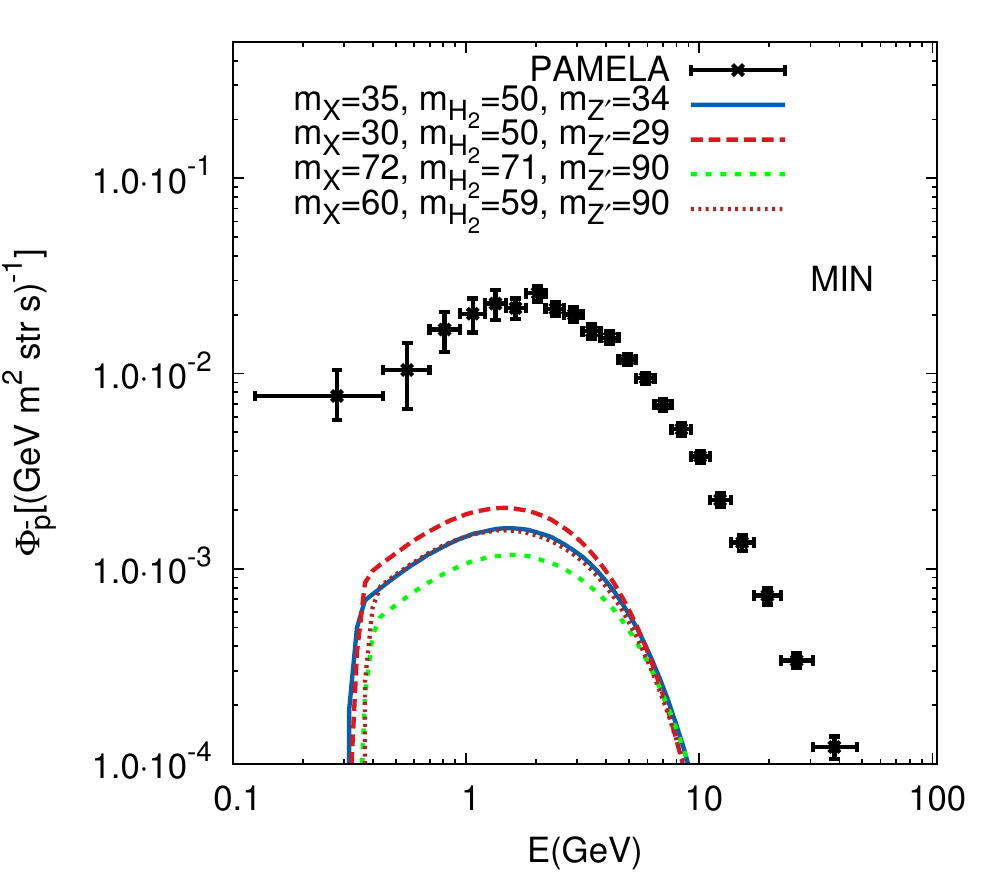}
\includegraphics[width=0.32\textwidth, height=0.32\textwidth]{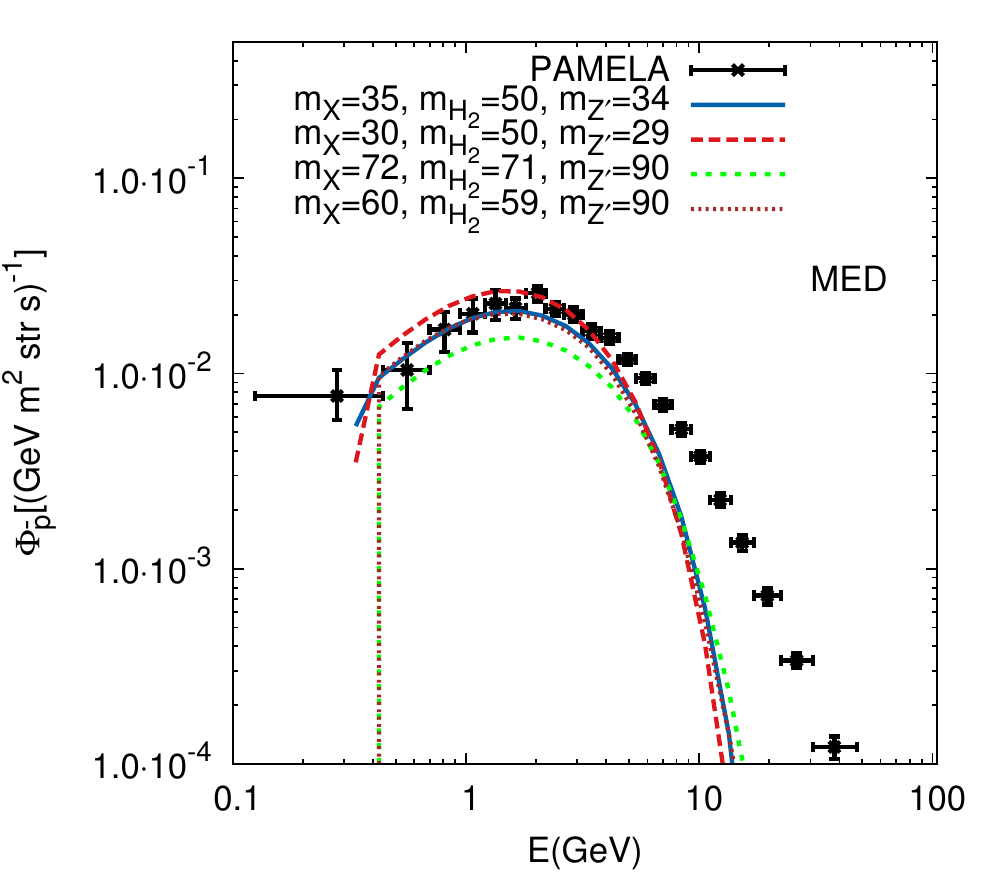}
\includegraphics[width=0.32\textwidth, height=0.32\textwidth]{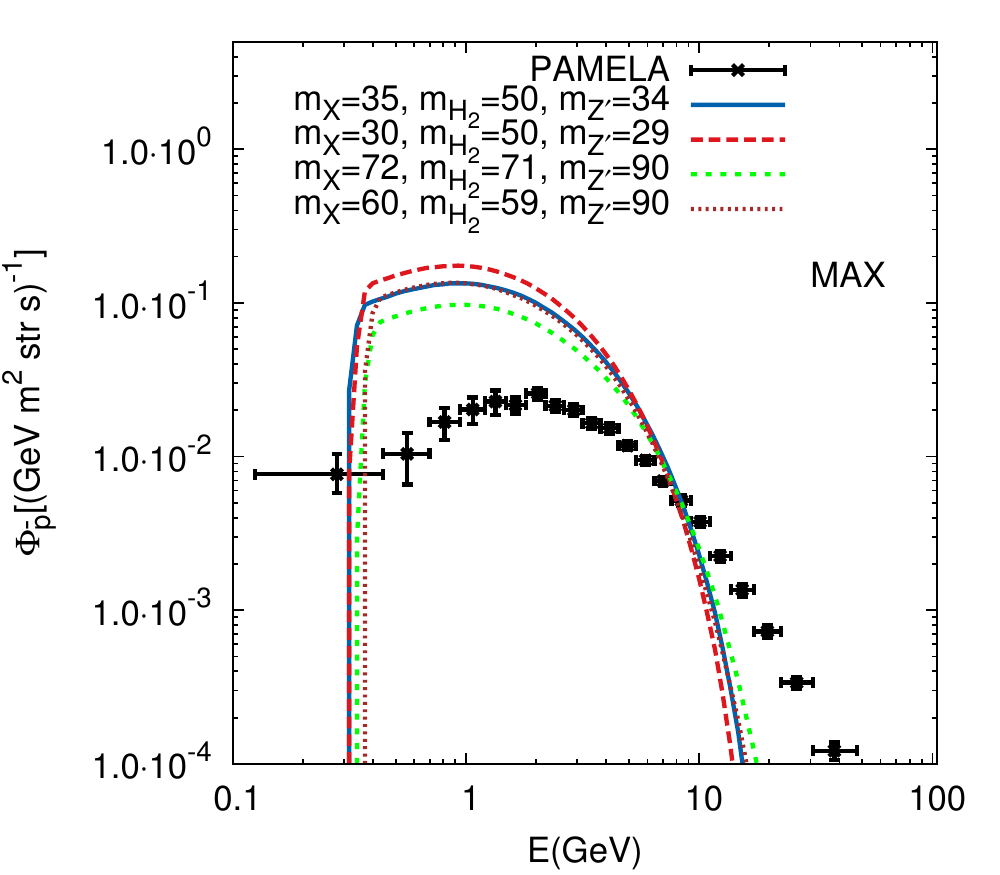}
\caption{Antiproton flux dependence on astrophysical parameters. 
From left to right, MIN, MED and MAX models are used respectively. See table.~\ref{tab:models} for model parameters.}
\label{fig:antiproton}
\end{figure}

As a concrete illustration, in Fig.~\ref{fig:antiproton} we show how the anti-proton flux can 
change for different astrophysical models with parameters shown in  Table~\ref{tab:models}, 
by solving Eq.~(\ref{eq:propa}) with \texttt{micrOMEGAs-3}~\cite{micromegas3}. 
As shown, because of these uncertainties,  there is still viable parameter space that is 
consistent with such constraints from cosmic ray spectra (see Ref.~\cite{Bringmann:2014lpa} 
for further detailed discussion including the constraints from radiowave).

\begin{table}[t]
  \centering
    \begin{tabular}{|c||c|c|c|c|}
    \hline 
 Model & $\delta$ & $K_0\;(\rm kpc^2/Myr)$ & $L\;(\rm kpc)$ & $V_C(\rm{km/s})$ \\\hline
  MIN & 0.85 & 0.0016 & 1 & 13.5 \\\hline
  MED & 0.7 & 0.0112 & 4 & 12 \\\hline
  MAX  & 0.46 & 0.0765 & 15 & 5 \\\hline 
    \end{tabular}
 \caption{Astrophysical models that are consistent with the B/C data~\cite{Maurin:2001sj, Donato:2003xg}. 
 $L$ is half of the thickness of diffusion zone for cosmic rays.  }
    \label{tab:models}
\end{table}

Neutrinos are also produced promptly from the above DM (semi-)annihilation with the absolute flux depending on the final states. Since the $\langle \sigma v\rangle _{\textrm{ann}}\sim 10^{-26}\textrm{cm}^3/$s, the  neutrino flux is about $3-4$ orders smaller than the current 
sensitivity or the limits from neutrino telescopes, such as IceCube or Super K. Unless there is 
a huge boost factor from astrophysics or other mechanisms, we expect that the neutrinos produced
from DM (semi-)annihilation can not be detected and therefore no meaningful constraints from 
neutrino flux measurements.

\section{Generalization to hidden sector DM models with local dark gauge symmetries}\label{sec:lesson}

From our discussions, it is clear that the gamma ray excess from the GC can be accommodated 
if there is a new particle analogous to the dark photon or dark Higgs boson with a suitable mass. This is in fact realized readily in a class of dark matter models with local dark 
gauge symmetry where DM is thermalized by singlet portals (including Higgs portal). 
In such models, there are almost always a SM singlet scalar $\phi$ from the dark sector
~\cite{Baek:2013dwa}, as well as a dark gauge boson $Z^{'}$, both of which couple to dark matter 
particle $X$, independent of the details of dark gauge symmetry or matter contents in the dark sector. 

Dark Higgs will modify the signal strengths of the observed Higgs boson in a simple and 
testable way: the signal strength will be reduced from ``1'' in a universal fashion, independent of 
production and decay channels of the SM Higgs boson~\cite{Baek:2011aa,Baek:2012se}.   
This is a generic aspect of renormalizable and unitary Higgs portal DM models with local dark 
gauge symmetries ~\cite{Baek:2013dwa,Baek:2013qwa,Ko:2014bka,localz3,Ko:2014uka} 
or without local dark gauge symmetries~\cite{Baek:2011aa,Baek:2012se}. 
Also nonstandard decays of the SM Higgs boson are possible, such as 
$H \rightarrow \phi \phi , Z^{'} Z^{'} , X X$, etc., including the invisible Higgs decay into a pair of DM particles (see Ref.~\cite{Chpoi:2013wga} for global analysis of Higgs signal strengths in the presence of light extra singlet scalar that mixes with the SM Higgs boson). 
Dark Higgs boson can also be helpful on improving vacuum stability up to Planck scale and 
the Higgs inflation (see Refs.~\cite{Baek:2012uj} and \cite{Ko:2014eia}, respectively, 
for examples).

\section{Conclusion}\label{sec:conclusion}
In conclusion, we discussed the galactic center $\gamma$-ray excess in the scalar DM model 
with local $Z_3$ symmetry that is the remnant of a spontaneously broken $U(1)_X$ dark 
gauge symmetry.  Due to the  newly-opened (semi-)annihilation channels involving dark Higgs 
and/or dark gauge boson, scalar DM as light as several ten GeV is still possible in the local 
$Z_3$ model, unlike in the global $Z_3$ model where such a DM mass range is not allowed by thermal relic density and direct detection constraints. In the local $Z_3$ scalar DM model, dark matter particles can (semi)-annihilate into the dark Higgs $H_2$s and/or dark photon $Z'$s  that immediately decay into light SM fermion pairs, such as  $b\bar{b}$ or $\tau \tau$, 
etc.. The $\gamma$ ray from these light fermions can fit the GeV scale $\gamma$-ray data reasonably well. Depending on the relative contributions of individual (semi-)annihilation channel, DM mass can vary from $30$GeV to 
$70$GeV, in a wide range of parameter space~\cite{localz3}. 

Finally we generalized this mechanism for the GeV scale $\gamma$-ray excess from the GC to DM models with local dark gauge symmetries, and argued that one can easily accommodate the 
GC $\gamma$ ray excess using the dark photon and/or dark Higgs boson which are generically 
present in such DM models.   
Particularly, dark Higgs can play more important role, since the dark 
gauge boson coupled to the SM fields through the $U(1)$ kinetic mixing at tree level
~\footnote{There could be kinetic mixing from higher dimensional operators, which would be 
however parametrically suppressed relative to the mixing from renormalizable Higgs portal interaction.}can be realized only in the Abelian dark gauge symmetry . 
In DM models with non-Abelian dark gauge symmetries where dark gauge bosons cannot have renormalizable couplings to the SM fields, one can invoke dark Higgs and Higgs portal interactions for the GC $\gamma$-ray excess. 
Dark Higgs couples more strongly to the heavier SM fermions such as $b\bar{b}$ or 
$\tau\tau$, and thus  its couplings are naturally flavor dependent as noticed in Ref.~\cite{kpt}. 
It is amusing to notice that DM models where DM is stabilized by the local dark gauge symmetry 
have a number of new fields which can be utilized for DM self-interaction \cite{localz3} or 
the $\gamma$-ray excess from the GC, depending on the mass spectra of these new particles. Dark Higgs boson can also mix with the SM Higgs boson and reduce the Higgs signal strengths from one in a universal manner, independent of production and decay channels. It also improves the stability of EW vacuum upto Planck scale and generate a larger tensor-to-scalar ratio $r \sim O(0.1)$~\cite{Ko:2014eia}. Therefore active searches for the dark Higgs at colliders are clearly warranted.

\begin{acknowledgments}
This work is supported in part by National Research Foundation of Korea (NRF) Research 
Grant 2012R1A2A1A01006053 (PK,YT), and by the NRF grant funded by the Korea 
government (MSIP) (No. 2009-0083526) through  Korea Neutrino Research Center 
at Seoul National University (PK). We acknowlege use of \texttt{Jaxodraw}\cite{jaxodraw} for drawing Feynman diagrams. 
\end{acknowledgments}

\end{document}